\documentclass[%
preprint,
 amsmath,amssymb,
 aps, physrev,
prx,
]{revtex4-2}

\usepackage{graphicx}
\graphicspath{figures{f/}}
\usepackage{dcolumn}
\usepackage{bm}
\usepackage{hyperref}

\usepackage{xcolor}

\begin{document}

\title{\textbf{\textbf{\textbf{N}}\textbf{\textbf{onmagnetic-magnetic }}\textbf{\textbf{Transitions}}\textbf{\textbf{ in Rutile RuO}}\textbf{\textsubscript{\textsubscript{\textbf{2}}}}} }%

\author{Yue-Fei Hou}
 \affiliation{
 School of Physics, Beijing Institute of Technology, Beijing 100081, China}
 \affiliation{
 Institute of Applied Physics and Computational Mathematics, Beijing 100088, China}
 \author{Siyuan Liu}
 \affiliation{
  School of Physics, Beijing Institute of Technology, Beijing 100081, China}
 \author{Wanxiang Feng}
 \affiliation{
  School of Physics, Beijing Institute of Technology, Beijing 100081, China}
\author{Jiajun Lu}
  \affiliation{
  School of Physics, Beijing Institute of Technology, Beijing 100081, China}
 \author{Xinfeng Chen}
 \affiliation{
 School of Physics, Beijing Institute of Technology, Beijing 100081, China}
\author{Gui-Bin Liu}
 \email{Contact author: gbliu@bit.edu.cn}
 \affiliation{
 School of Physics, Beijing Institute of Technology, Beijing 100081, China}
\author{Ping Zhang}
\email{Contact author: zhang\_ping@iapcm.ac.cn}
\affiliation{
 Institute of Applied Physics and Computational Mathematics, Beijing 100088, China}
 
\date{\today}

\begin{abstract}

Rutile RuO$_2$ has recently attracted great interest, as its magnetic ground state remains controversial. Experimental studies have reported either nonmagnetic (NM) or altermagnetic (AM) ground states in different crystalline samples of RuO$_2$, highlighting the need for a reasonable explanation to resolve this contradiction. In this study, density functional theory calculations are performed to reveal the correlation-sensitive and strain-dependent magnetism of bulk RuO$_2$. On one hand, multiple AM phases with different magnitudes of the spin magnetic moment are identified in the Hubbard parameter space for RuO$_2$. On the other hand, when appropriate strains that significantly change the crystal cell volume are applied, the ground state of RuO$_2$ can undergo transitions between the NM state (with no spin splitting) and the magnetic states (with spin splitting in the band structure). These findings not only demonstrate intriguing physics in 4\textit{\textit{d}}-electron-correlated RuO$_2$, but also retain its potential for spintronic applications.

\end{abstract}

\maketitle
 

\section{\label{sec:level1}Introduction}

Rutile RuO$_2$ has been recognized as a Pauli paramagnet since the last century \cite{Ryden1970a,Ryden1970b,Graebner1976}, before itinerant antiferromagnetism was first evidenced by polarized neutron diffraction \cite{Berlijn2017}. Later, resonant X-ray scattering \cite{Zhu2019,Lovesey2022}, angle-resolved photoelectron spectroscopy \cite{Lin2024,Fedchenko2024}, X-ray magnetic circular/linear dichroism \cite{Fedchenko2024,Lytvynenko2026,Zhang2025,He2025}, spin-torque ferromagnetic resonance \cite{Guo2024}, magneto-optic Kerr effect measurements \cite{Jeong2026a,Zhang2025b} and planar Hall effect measurements \cite{Song2024} also evidenced its antiferromagnetism (AFM), or rather the so-called altermagnetism (AM) \cite{Smejkal2022a,Smejkal2022b}—with spin splitting appearing in the band structure. However, abundant experiments simultaneously support the nonmagnetic (NM) nature of rutile RuO$_2$, such as two other spin-resolved photoelectron spectroscopy studies \cite{Liu2024,Osumi2026}, muon spin resonance \cite{Hiraishi2024,Kessler2024}, another polarized neutron diffraction study \cite{Kiefer2025}, infrared spectroscopy \cite{Wenzel2025}, nuclear quadrupole resonance measurements \cite{Song2025}, Mössbauer spectroscopy \cite{Yumnam2025}, another X-ray linear dichroism study \cite{Wang2026}, and transport measurements \cite{Wang2026b}. In view of these two opposite but reliable viewpoints, the true magnetic ground state of RuO$_2$ and reasonable explanations require further investigation.

Recently, to address this unresolved problem, an increasing number of studies have indicated that the specific conditions of different RuO$_2$ crystalline samples may lead to different types of magnetism. Although not universally observed, magnetic features are more likely to be found in thin-film samples \cite{Fedchenko2024,Lytvynenko2026,Zhang2025,He2025,Guo2024,Jeong2026a,Zhang2025b,Jeong2026b,Lee2026,Liang2022,Ho2025,Brahimi2024,Akashdeep2026,Howzen2026,Wickramaratne2026,}, while NM features typically appear in bulk samples \cite{Osumi2026,Hiraishi2024,Kessler2024,Kiefer2025,Wenzel2025,Song2025,Yumnam2025,Peng2025,Occhialini2026,Pawula2024,Mukuda1999,Hariki2024}. On one hand, this phenomenon suggests that size effects \cite{Jeong2026a,Zhang2025b,Liang2022,Ho2025,Brahimi2024,Jeong2025}, surface effects \cite{Lytvynenko2026,Zhang2025b,Yumnam2025,Liang2022,Ho2025,Brahimi2024,Akashdeep2026,Howzen2026,Torun2013}, crystal defects \cite{Wang2026,Jiang2026,Smolyanyuk2024}, and epitaxial strains \cite{Fedchenko2024,Lytvynenko2026,Jeong2026a,Zhang2025b,Song2024,Wang2026,Jeong2026b,Lee2026,Jeong2025,Jiang2026,Gregory2022,Wickramaratne2026,gqhb-2h45} may play key roles in inducing the magnetic states, reversing the possibility of AM order in RuO$_2$ films. On the other hand, the mechanism underlying both the NM and magnetic characters in bulk RuO$_2$ remains intriguing topics. Although the origin of magnetic transitions in bulk RuO$_2$ has been attributed to Fermi surface instability by previous studies \cite{Qian2025,Ahn2019}, the present explanations lack reasonably quantitative analyses regarding the magnetic-transition criterion and the magnitude of the magnetic moment. Providing such discussions will not only deepen the understanding of condensed-matter magnetism, but also promote the practical application of altermagnetism in RuO$_2$ bulk.

In this work, we perform density functional theory (DFT) calculations to reveal the nonmagnetic–magnetic transitions in rutile RuO$_2$ bulk. We once again show that the electronic correlation of 4\textit{\textit{d}} electrons is essential for stabilizing magnetic moments in RuO$_2$. Due to the sensitivity to the 4$d$-electron correlation strength, the DFT-calculated AM states are found to be not unique within a reasonable Hubbard-parameter range, exhibiting significantly different magnitudes of the spin magnetic moment. By imposing strains on the ideal RuO$_2$ crystal, we find that the magnetic ground state can undergo transitions between the NM state and the magnetic ordered states. Our calculation results show that whether an imposed strain mode induces a magnetic transition is directly related to the resulting variation in crystal cell volume. It is also found that both 4$d$-electron correlation and the electronic density of states (DOS) of the NM state influence the stability of the magnetic state of RuO$_2$, which attributes the origin of magnetism in RuO$_2$ to a generalized Stoner mechanism. In the end, the longitudinal conductivities of bulk RuO$_2$ under continuous strain are simulated. The results show that conductivity measurement is a straightforward method to detect the NM-magnetic transitions in RuO$_2$. Our findings provide additional insights for a reasonable understanding of the experiments and suggest a straightforward mechanism for tuning the magnetic moment in itinerant-electron systems with considerable electronic correlation.

\section{\label{sec:level1}Results and analysis }
\subsection{\label{sec:level2}Electronic structures of the NM and AM states}
\begin{figure}[htbp]
	\centering
	\includegraphics[width=1.0\textwidth]{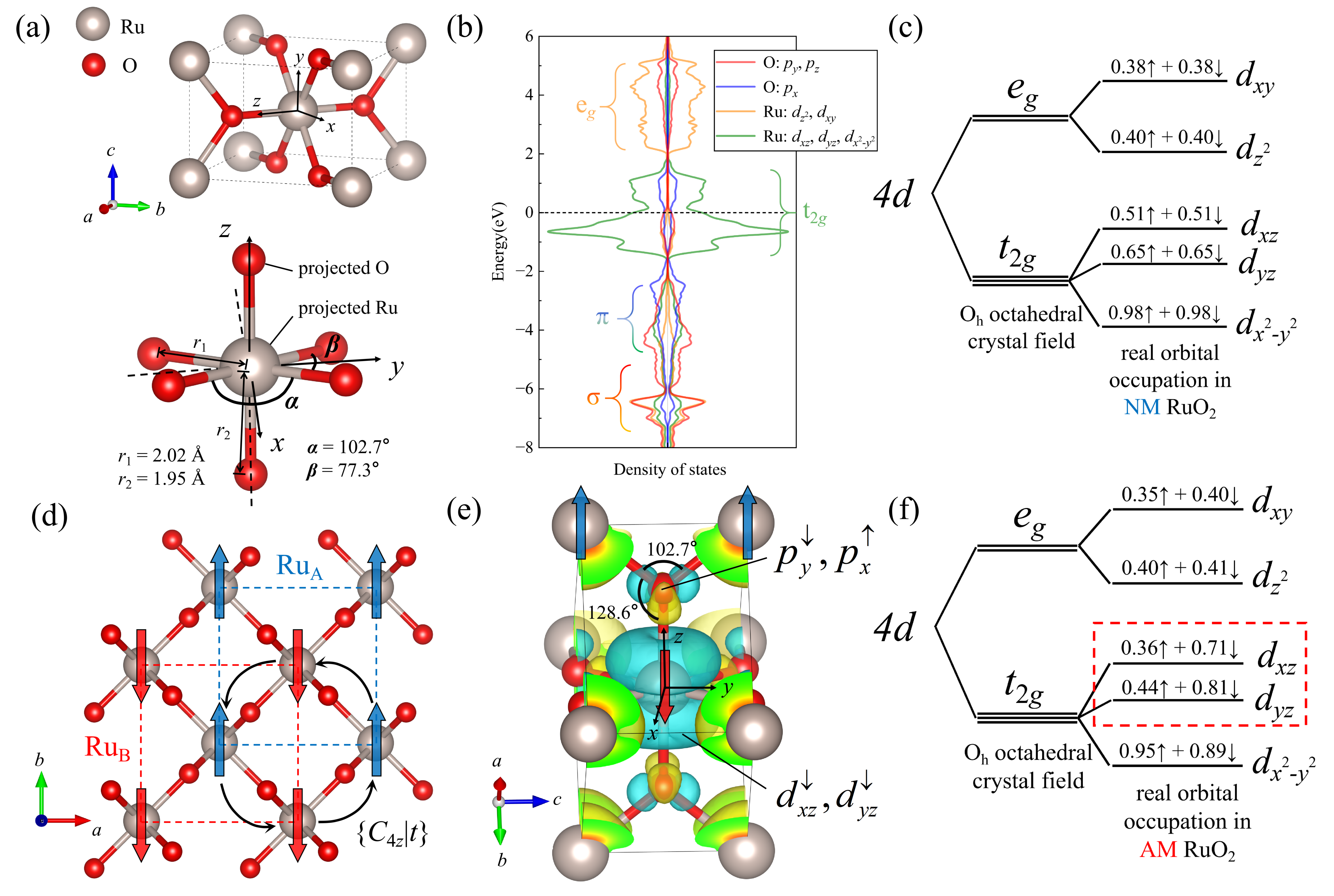}
	\caption{(a) The atomic structure of rutile RuO$_2$. The local Cartesian coordinate system is set on the central Ru atom in the unit cell. \textit{\textit{r}}\textsubscript{\textsubscript{1}} and \textit{\textit{r}}\textsubscript{\textsubscript{2}} are the Ru-O bond lengths in the \textit{\textit{x-y}} plane and along \textit{\textit{z}} direction, respectively. \textit{\textit{$\alpha$}} and \textit{\textit{$\beta$}} are the O-Ru-O bond angles in the \textit{\textit{x-y}} plane. The Ru and O atoms used for orbital projection are also marked by the black lines. (b) The orbital-projected electronic DOS of RuO$_2$ based on the local coordinate system introduced in (a). (c) The spin occupation on the 4\textit{\textit{d}} orbitals of the NM state suggested by DFT calculations. (d) The collinear AM structure of RuO$_2$. The two Ru sublattices are represented by Ru\textsubscript{\textsubscript{A}} and Ru\textsubscript{\textsubscript{B}}, respectively. The blue and red arrows represent different spin magnetic moments for each Ru sublattice. (e) The net spin density of the AM state of RuO$_2$. The yellow and blue isosurfaces represent spin up and spin down densities, respectively. The isosurfaces correspond to the spin density of 3×10\textsuperscript{\textsuperscript{-3}} Å\textsuperscript{\textsuperscript{-}}\textsuperscript{\textsuperscript{3}}. The two black lines mark the main contributing orbitals to the spin densities of Ru and O. The other orbitals’ contributions are negligible, as they are almost an order of magnitude smaller than those of the mainly contributing orbitals. (f) The spin occupation on the 4\textit{\textit{d}} orbitals of the AM state suggested by DFT calculations. The red dashed box highlights the main contributing orbitals to the spin magnetic moment of Ru.
	}
	\label{fig1}
\end{figure}

The electronic structures of the NM state and the magnetic ordered states have been extensively studied by previous DFT calculations. Here, we first recalculate the DFT electronic structures and then show the orbital composition of the spin magnetic moments in RuO$_2$. Detailed computational parameters and methods are introduced in Section A of the Supplementary Material (SM) \cite{SM}. As shown in Fig. \ref{fig1}(a), the local Cartesian coordinate system for orbital projection is marked in the crystal structure. Around the projected Ru atom in the center, there is a distorted RuO$_6$ octahedral structure, with the Ru-O bond length along the \textit{\textit{z}} direction slightly shortened and the O-Ru-O bond angles in the \textit{\textit{x–y}} plane significantly changed. The point group symmetry of the crystal is thus lowered from O\textsubscript{\textsubscript{h}} to D\textsubscript{\textsubscript{2h}} \cite{Occhialini2021}.

In the NM state of RuO$_2$, the two spins equally and independently fill the band structure. The orbital-projected DOS from DFT calculations is shown in Fig. \ref{fig1}(b). Although the RuO$_6$ octahedron is distorted, the \textit{\textit{e}}\textit{\textsubscript{\textsubscript{\textit{g}}}} - \textit{\textit{t}}\textsubscript{\textsubscript{2}}\textit{\textsubscript{\textsubscript{\textit{g}}}} crystal field (CF) splitting under O$_h$ symmetry can still be roughly recognized, as suggested by inelastic X-ray scattering \cite{Occhialini2021}. Fig. \ref{fig1}(c) presents the DFT-calculated Ru-4\textit{\textit{d}} spin occupation for the NM state. The low-energy \textit{\textit{d}}\textit{\textsubscript{\textsubscript{\textit{x}}}}\textsubscript{\textsubscript{2}}\textit{\textsubscript{\textsubscript{\textit{-y}}}}\textsubscript{\textsubscript{2}} orbital is nearly fully filled by two opposite spins, just as previous studies have suggested \cite{Berlijn2017,Jeong2026a,Smolyanyuk2024,Ahn2019,Occhialini2021}. The other four orbitals are partially filled, and their occupation numbers can be reasonably explained by their relative CF energies in RuO$_2$\textsubscript{\textsubscript{.}}

The AM ordered state of RuO$_2$ is strongly suggested due to the constraints of crystal symmetry \cite{Smejkal2022a,Smejkal2022b}. The two spin sublattices are related by a rotation-plus-translation operation \{\textit{\textit{C}}\textsubscript{\textsubscript{4z}}\textbar \textbf{\textit{\textit{\textbf{t}}}})\}, as shown in Fig. \ref{fig1}(d). Owing to the equivalence of the two sublattices, the local magnetic moment of each Ru atom has exactly the same distribution of net spin density under the local coordinate system. In Fig. \ref{fig1}(e), the net spin density of the AM RuO$_2$ calculated by DFT is shown. At first sight, the gourd-shaped net spin density of Ru in RuO$_2$ is mainly composed of \textit{\textit{d}}\textit{\textsubscript{\textsubscript{\textit{xz}}}} and \textit{\textit{d}}\textit{\textsubscript{\textsubscript{\textit{yz}}}} orbitals’ contributions. Indeed, the projected net spin for \textit{\textit{d}}\textit{\textsubscript{\textsubscript{\textit{xz}}}} and \textit{\textit{d}}\textit{\textsubscript{\textsubscript{\textit{yz}}}} orbitals are 0.35 and 0.37 spin/Ru, respectively, with the total spin magnetic moment of about 0.72 $\mu$\textsubscript{\textsubscript{B}}, as the two orbitals hold the same net spin. The net spin density of O is mainly composed of contributions from \textit{\textit{p}}\textit{\textsubscript{\textsubscript{\textit{y}}}} and \textit{\textit{p}}\textit{\textsubscript{\textsubscript{\textit{x}}}} orbitals. The projected net spin are about 0.01 spin/O for both \textit{\textit{p}}\textit{\textsubscript{\textsubscript{\textit{y}}}} and \textit{\textit{p}}\textit{\textsubscript{\textsubscript{\textit{x}}}} orbitals, while the total spin moment of O cancels out because the two orbitals hold opposite net spins. Nonetheless, the O ligand should play a key role in the superexchange interactions between the magnetic Ru atoms due to its unfulfilled \textit{\textit{p}} orbitals. Our DFT calculations suggest a spin-parallel order along the \textit{\textit{c}} axis of the crystal and a spin-antiparallel order between the two sites of the two sublattices. This is consistent with previous DFT studies \cite{Berlijn2017,Jeong2026a,Lee2026,Jeong2025,Smolyanyuk2024,Occhialini2021}. This AM ground state is also well explained by the Goodenough rules regarding orbital occupation and bond angles \cite{Goodenough1960}. 

In Fig. \ref{fig1}(f), the DFT-calculated Ru-4\textit{\textit{d}} spin occupation for the AM state is shown. Compared with the results for the NM state shown in Fig. \ref{fig1}(c), no significant CF-related 4$d$ inter-orbital transitions are observed to explain the formation of the local magnetic moments. This is not similar to the situation in marcasite FeTe$_2$ \cite{Hou2025}, where the formation of spin magnetic moments requires the 3\textit{\textit{d}} inter-orbital transitions from the NM state. The spin polarization appears independently on the \textit{\textit{d}}\textit{\textsubscript{\textsubscript{\textit{xz}}}} and \textit{\textit{d}}\textit{\textsubscript{\textsubscript{\textit{yz}}}} orbitals, implying a physical picture that combines both localized- and itinerant-electron magnetism in RuO$_2$.

\subsection{\label{sec:level2}4$d$-electron-correlation-stabilized spin magnetic moment}
\begin{figure}[htbp]
	\centering
	\includegraphics[width=0.5\textwidth]{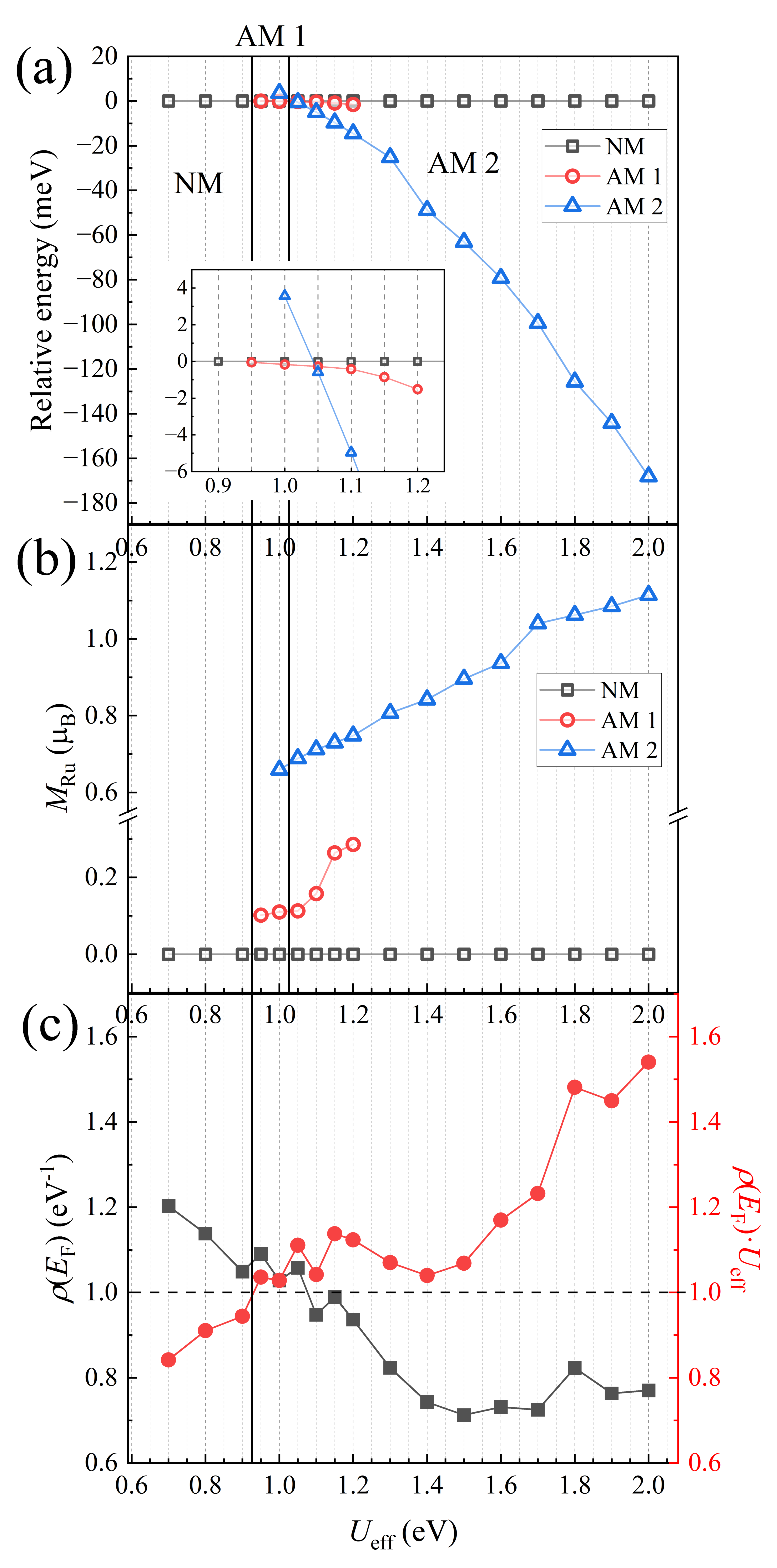}
	\caption{The physical properties of RuO$_2$ calculated with different values of \textit{\textit{U}}\textsubscript{\textsubscript{eff}}. (a) The relative total energies of the NM, AM 1, and AM 2 states. The black vertical lines divide the \textit{\textit{U}}\textsubscript{\textsubscript{eff}} parameter space into three different magnetic phases. (b) The Ru spin magnetic moments of the NM, AM 1, and AM 2 states. (c) The Fermi level’s DOS (black squares) of the NM state and the corresponding \textit{\textit{$\rho$}}(\textit{\textit{E}}\textsubscript{\textsubscript{F}}) $\cdot$ \textit{\textit{U}}\textsubscript{\textsubscript{eff }}values (red spheres).}
	\label{fig2}
\end{figure}

Previous studies have shown that standard DFT calculations fail to generate the AM ordered state of RuO$_2$ \cite{Lytvynenko2026,Jeong2026a,Wang2026,Lee2026,Liang2022,Jeong2025,Smolyanyuk2024}, suggesting that a reasonable correction for 4\textit{\textit{d}}-electron correlation is necessary. We reexamine the effects of the 4\textit{\textit{d}} on-site Coulomb parameter \textit{\textit{U}}\textsubscript{\textsubscript{eff}} from the Hubbard model \cite{Dudarev1998} used in our calculations on the physical properties, as shown in Fig. \ref{fig2}. By using different values of \textit{\textit{U}}\textsubscript{\textsubscript{eff}} in the range of 0.7 eV to 2.0 eV, the NM state and two AM states (namely AM 1 and AM 2) are identified. In Fig. \ref{fig2}(a), the total energies of the AM states relative to the NM state are provided. When \textit{\textit{U}}\textsubscript{\textsubscript{eff}} is less than 0.9 eV, only the NM state of RuO$_2$ is stable. The AM 1 and AM 2 states begin to be the ground states when \textit{\textit{U}}\textsubscript{\textsubscript{eff}} reaches 0.95 eV and 1.05 eV, respectively. Note that the AM states are only stable in certain ranges of \textit{\textit{U}}\textsubscript{\textsubscript{eff}}. The most notable NM-AM transition occurs at \textit{\textit{U}}\textsubscript{\textsubscript{eff}} = 0.95 eV. One can simply determine the AM region in the \textit{\textit{U}}\textsubscript{\textsubscript{eff}} parameter space; however, the multi-AM-phase feature is proposed here for RuO$_2$ for the first time, reflecting the sensitivity of magnetism to the 4\textit{\textit{d}} electron correlation strength in RuO$_2$.

The magnetic moments of Ru can be used to define the order parameter, such as the Néel vector, and are therefore calculated and shown in Fig. \ref{fig2}(b). As expected, the magnetic moments of Ru for the AM states increase with increasing \textit{\textit{U}}\textsubscript{\textsubscript{eff}}. In the AM 1 phase, where \textit{\textit{U}}\textsubscript{\textsubscript{eff}} ranges from 0.95 eV to 1.0 eV, the magnetic moment of Ru is small, approximately 0.1 $\mu$\textsubscript{\textsubscript{B}}. This provides a insight into the tiny magnetic responses observed in experiments \cite{Berlijn2017,Zhu2019}. In the AM 2 phases, where \textit{\textit{U}}\textsubscript{\textsubscript{eff}} is larger than 1.05 eV, the magnetic moment of Ru is significant and can exceed 0.7 $\mu$\textsubscript{\textsubscript{B}}. Given that no large magnetic moment of Ru has been directly observed by experiments, we estimate that a reasonable \textit{\textit{U}}\textsubscript{\textsubscript{eff}} shouldn't exceed 1.6 eV in RuO$_2$. More detailed computational results for the NM and two AM states, which help to understand their electronic and magnetic properties, are provided in Section B of the SM \cite{SM}.

To understand the mechanism of the NM-AM transition under varying \textit{\textit{U}}\textsubscript{\textsubscript{eff}}, we examine the stability of the magnetic state based on Stoner’s criterion: \textit{\textit{$\rho$}}(\textit{\textit{E}}\textsubscript{\textsubscript{F}})$\cdot$\textit{\textit{I}}\textsubscript{\textsubscript{eff}} $>$ 1 \cite{Stoner1938}, where \textit{\textit{$\rho$}}(\textit{\textit{E}}\textsubscript{\textsubscript{F}}) is the NM-state DOS at the Fermi level for each spin and \textit{\textit{I}}\textsubscript{\textsubscript{eff}} is the effective Stoner parameter. For the correlation-driven spin polarization in RuO$_2$, we simply replace \textit{\textit{I}}\textsubscript{\textsubscript{eff}} with \textit{\textit{U}}\textsubscript{\textsubscript{eff}}. Since there are two Ru sites in the unit cell of RuO$_2$, \textit{\textit{$\rho$}}(\textit{\textit{E}}\textsubscript{\textsubscript{F}}) for each Ru sublattice, along with the corresponding values of \textit{\textit{$\rho$}}(\textit{\textit{E}}\textsubscript{\textsubscript{F}})$\cdot$\textit{\textit{U}}\textsubscript{\textsubscript{eff}}, are shown in Fig. \ref{fig2}(c). Although DFT calculations indicate that \textit{\textit{$\rho$}}(\textit{\textit{E}}\textsubscript{\textsubscript{F}}) decreases with increasing \textit{\textit{U}}\textsubscript{\textsubscript{eff}}, the value of \textit{\textit{$\rho$}}(\textit{\textit{E}}\textsubscript{\textsubscript{F}})$\cdot$\textit{\textit{U}}\textsubscript{\textsubscript{eff}} indeed increases due to the influence of the \textit{\textit{U}}\textsubscript{\textsubscript{eff}} factor. Moreover, \textit{\textit{$\rho$}}(\textit{\textit{E}}\textsubscript{\textsubscript{F}})$\cdot$\textit{\textit{U}}\textsubscript{\textsubscript{eff}} becomes greater than 1 exactly when \textit{\textit{U}}\textsubscript{\textsubscript{eff}} exceeds 0.95 eV. This threshold is highly consistent with the point that spin magnetic moment emerges as shown in Fig. \ref{fig2}(b). Meanwhile, the approximation of using \textit{\textit{U}}\textsubscript{\textsubscript{eff}} as the effective Stoner parameter is proved suitable for RuO$_2$ in reproducing the NM-AM transition as shown in Fig. \ref{fig2}(a). In fact, achieving \textit{\textit{$\rho$}}(\textit{\textit{E}}\textsubscript{\textsubscript{F}})$\cdot$\textit{\textit{I}}\textsubscript{\textsubscript{eff}} $>$ 1 by imposing a large \textit{\textit{U}}\textsubscript{\textsubscript{eff}} theoretically leads to a ferromagnetic order. Our calculated \textit{\textit{$\rho$}}(\textit{\textit{E}}\textsubscript{\textsubscript{F}})$\cdot$\textit{\textit{U}}\textsubscript{\textsubscript{eff}} is exactly for each ferromagnetic Ru sublattice in the unit cell. Our calculations show the effectiveness of Stoner's criterion in predicting the spin magnetic moment, even in spin-antiparalell AM systems.

\subsection{\label{sec:level2}Strain-induced NM-magnetic transitions}
\begin{figure}[htbp]
	\centering
	\includegraphics[width=1.0\textwidth]{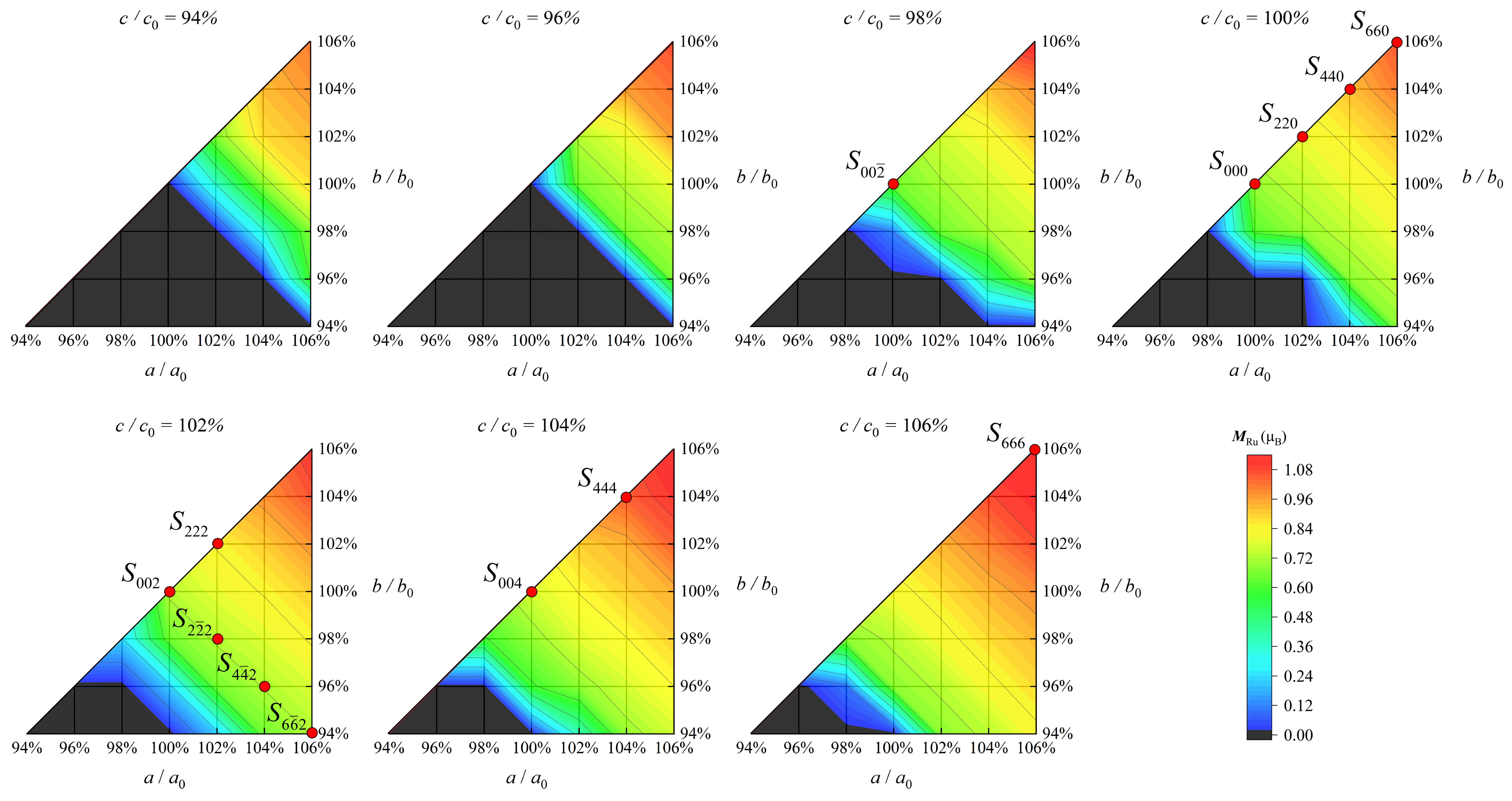}
	\caption{The NM-magnetic phase diagram of strained RuO$_2$ bulk. The magnitude of spin magnetic moment \textit{\textit{M}}\textsubscript{\textsubscript{Ru}} for different strained states are represented by the chroma. \textit{\textit{a}}, \textit{\textit{b}}, and \textit{\textit{c }}axes of RuO$_2$ are adjusted in the range 94\% - 106\% of their experimental values ($a_0$ = $b_0$ = 4.49 Å, $c_0$ = 3.11 Å) \cite{Zhu2019}, with the step of 2\% to sample the whole strained-state space. The intersection points of the grid lines mark the actually calculated points. The intermediate areas between calculated points are completed by linear interpolation.}
	\label{fig3}
\end{figure}

Based on experimental results, it is inferred that epitaxial strains can lead to the magnetism of RuO$_2$ films \cite{Fedchenko2024,Lytvynenko2026,Jeong2026a,Zhang2025b,Song2024,Wang2026,Jeong2026b,Lee2026,Jeong2025,Jiang2026,Qian2025}. In this study, we explore magnetic transitions induced by strain only in bulk RuO$_2$, where surface and defect effects that could influence magnetism are excluded. Fig. \ref{fig3} shows the calculated spin magnetic moments of Ru for strained RuO$_2$, with the \textit{\textit{a}}, \textit{\textit{b}}, and \textit{\textit{c}} crystal axes independently compressed or stretched. In this phase diagram, the NM-to-magnetic transition can be easily identified by the black-to-colored boundary. In Section C of the SM \cite{SM}, another two calculated phase diagrams are given with different values of \textit{\textit{U}}\textsubscript{\textsubscript{eff}}. For convenience, the strained state of each calculated point in the phase diagram is labeled as \textit{\textit{S}}\textit{\textsubscript{\textsubscript{\textit{ijk}}}}, where the numbers \textit{\textit{i}}, \textit{\textit{j}}, and \textit{\textit{k}} represent the percentage changes relative to the experimental lattice parameters, respectively. For example, \textit{\textit{S}}\textsubscript{\textsubscript{2$\bar{2}$0}} denotes a strained state of RuO$_2$ bulk in which the \textit{\textit{a}} axis is stretched by 2\%, the \textit{\textit{b}} axis is compressed by 2\%, and the \textit{\textit{c}} axis remains unchanged.

In general, Fig. \ref{fig3} shows that the NM state of RuO$_2$ bulk is stabilized by lattice compression, while the magnetic state is stabilized by lattice expansion. This is also supported by a previous DFT study \cite{Qian2025}. Interestingly, the magnitude of the spin magnetic moment exhibits an almost linear relationship with the cell volume of RuO$_2$, particularly in the significantly stretched region of the lattice. This suggests a simple and straightforward mechanism governing the magnetic moment in strained RuO$_2$.

\begin{figure}[htbp]
	\centering
	\includegraphics[width=1.0\textwidth]{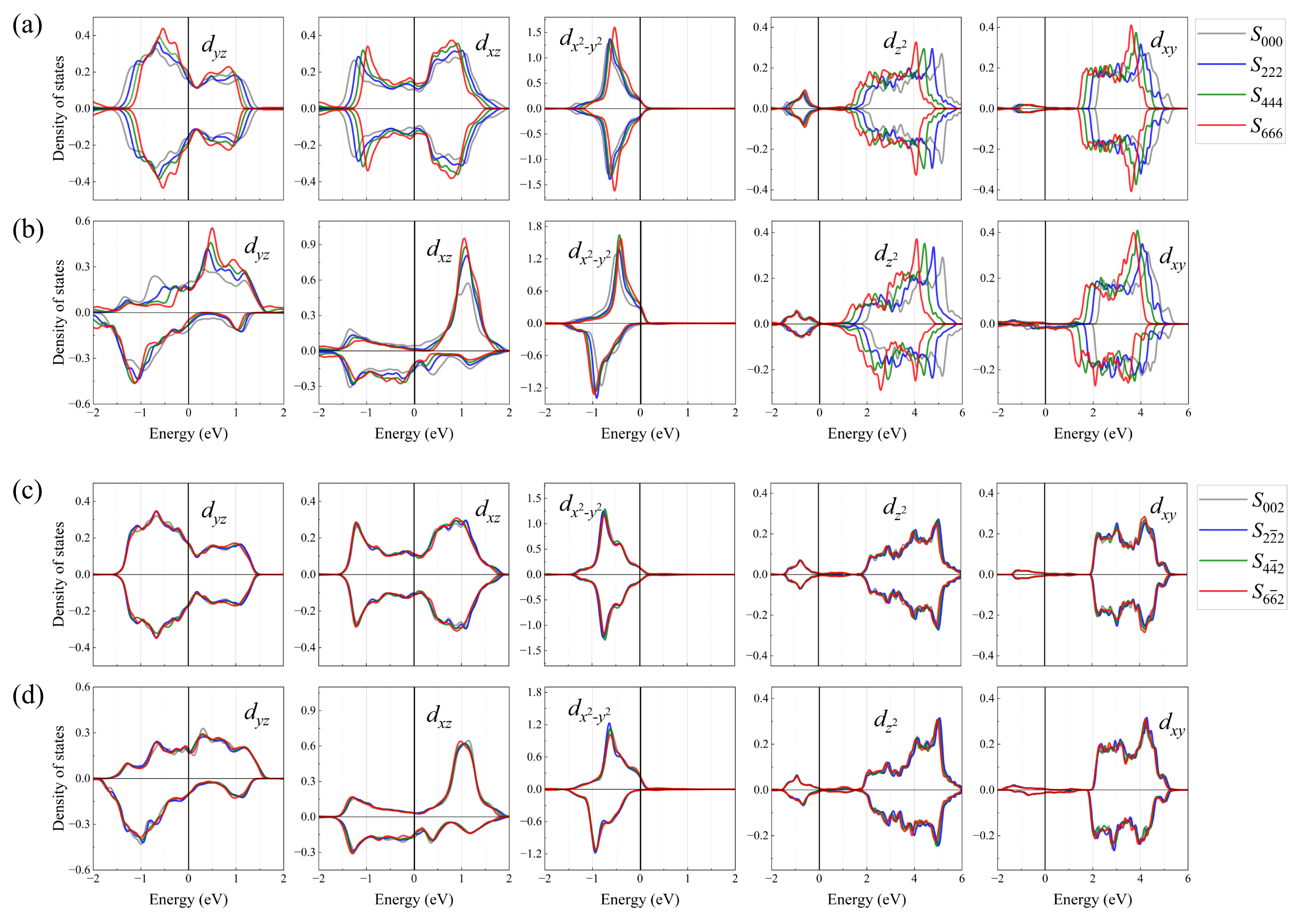}
	\caption{The orbital-projected DOSs of Ru-4$d$ orbitals for different strained states. Spin-up and Spin-down DOSs are represented by the positive and negative values, respectively. (a): The NM states and (b): the magnetic states of RuO$_2$ with the black, blue, green and red lines representing the DOSs of\textit{\textit{ S}}\textsubscript{\textsubscript{000}}, \textit{\textit{S}}\textsubscript{\textsubscript{222}}, \textit{\textit{S}}\textsubscript{\textsubscript{444}} and \textit{\textit{S}}\textsubscript{\textsubscript{666}} strained states, respectively. (c): The NM states and (d): the magnetic states of RuO$_2$ with the black, blue, green and red lines representing the DOSs of\textit{\textit{ S}}\textsubscript{\textsubscript{002}}, \textit{\textit{S}}\textsubscript{\textsubscript{2$\bar{2}$2}}, \textit{\textit{S}}\textsubscript{\textsubscript{4$\bar{4}$2}} and \textit{\textit{S}}\textsubscript{\textsubscript{6$\bar{6}$2}} strained states, respectively. }
	\label{fig4}
\end{figure}

To understand the enhancement of the magnetic moment under lattice expansion, the projected densities of states (PDOSs) on the Ru-4\textit{\textit{d}} orbitals are calculated and presented in Fig. \ref{fig4}. To begin with, the NM state is always instructive for theoretically studying the magnetic state. As the lattice isotropically expands from \textit{\textit{S}}\textsubscript{\textsubscript{000}} to \textit{\textit{S}}\textsubscript{\textsubscript{666}}, the PDOSs of the \textit{\textit{d}}\textit{\textsubscript{\textsubscript{\textit{yz}}}} and \textit{\textit{d}}\textit{\textsubscript{\textsubscript{\textit{xz}}}} orbitals in the NM state become narrower, as shown in Fig. \ref{fig4}(a). In other words, the localization of 4\textit{\textit{d}} electrons is strengthened with the increasing cell volume. According to our analysis in Section D of the SM \cite{SM}, this change in the geometric structures of the 4$d$-orbital PDOSs produces more electronic states near the Fermi level. For a finite Hubbard type splitting of the spin band structure in RuO$_2$, a larger spin magnetic moment on Ru can be expected in a more expanded lattice. As for the fully filled \textit{\textit{d}}\textit{\textsubscript{\textsubscript{\textit{x}}}}\textsubscript{\textsubscript{2}}\textsubscript{\textsubscript{-}}\textit{\textsubscript{\textsubscript{\textit{y}}}}\textsubscript{\textsubscript{2}} orbital and the partially filled \textit{\textit{d}}\textsubscript{\textsubscript{z}}\textsubscript{\textsubscript{2}} and \textit{\textit{d}}\textit{\textsubscript{\textsubscript{\textit{xy}}}} orbitals, although their PDOS distributions are also narrowed, the nearly zero PDOSs immediately above the Fermi level prevent these orbitals from contributing significantly to the spin magnetic moment. This explanation theoretically supports the DFT-calculated spin density shown in Fig. \ref{fig1}(e). As illustrated in Fig. \ref{fig4}(b), the DFT-calculated 4$d$-orbital PDOSs of the magnetic states indeed confirm the enhancement of the spin magnetic moment under lattice expansion from the perspective of first-principles. When equal amounts of tensile strain and compressive strain are imposed on $a$ and $b$ axes, respectively, there accompanies a tetragonal-to-orthorhombic distortion for the RuO$_2$ lattice but with an unchanged cell volume. Interestingly, the NM-state PDOSs of 4$d$ orbitals are thus nearly unchanged within this distortion, as shown in Fig. \ref{fig4}(c). Likewise, as shown in Fig. \ref{fig4}(d), the DFT-calculated 4$d$-orbital PDOSs of their corresponding magnetic states also give the unchanged spin magnetic moment. In this context, the cell-volume-controlled 4$d$-orbital PDOSs of the NM state is suggested to be a key physical quantity for predicting the magnetic moment in RuO$_2$ under a rigid-band approximation.

In the magnetic-state region, the AM ordered state is favored only if the \{C\textsubscript{\textsubscript{4}}\textit{\textsubscript{\textsubscript{\textit{z}}}}$|$\textit{\textit{t}}\} symmetry operation relating the two spin sublattices is maintained, i.e., for \textit{\textit{S}}\textit{\textsubscript{\textsubscript{\textit{ijk}}}} with \textit{\textit{i}} = \textit{\textit{j}}, such as \textit{\textit{S}}\textsubscript{\textsubscript{222}}, \textit{\textit{S}}\textsubscript{\textsubscript{004}}, or \textit{\textit{S}}\textsubscript{\textsubscript{660}}. For strained states with \textit{\textit{i}} $\ne$ \textit{\textit{j}}, the \{C\textsubscript{\textsubscript{4}}\textit{\textsubscript{\textsubscript{\textit{z}}}}$|$\textit{\textit{t}}\} operation is broken, and the ground state consequently becomes a compensated ferrimagnetic (FIM) ordered state, with spin splitting extending across the entire Brillouin zone. Nevertheless, controlling the degree of spin splitting is of interest for spintronics applications.

\begin{figure}[htbp]
	\centering
	\includegraphics[width=1.0\textwidth]{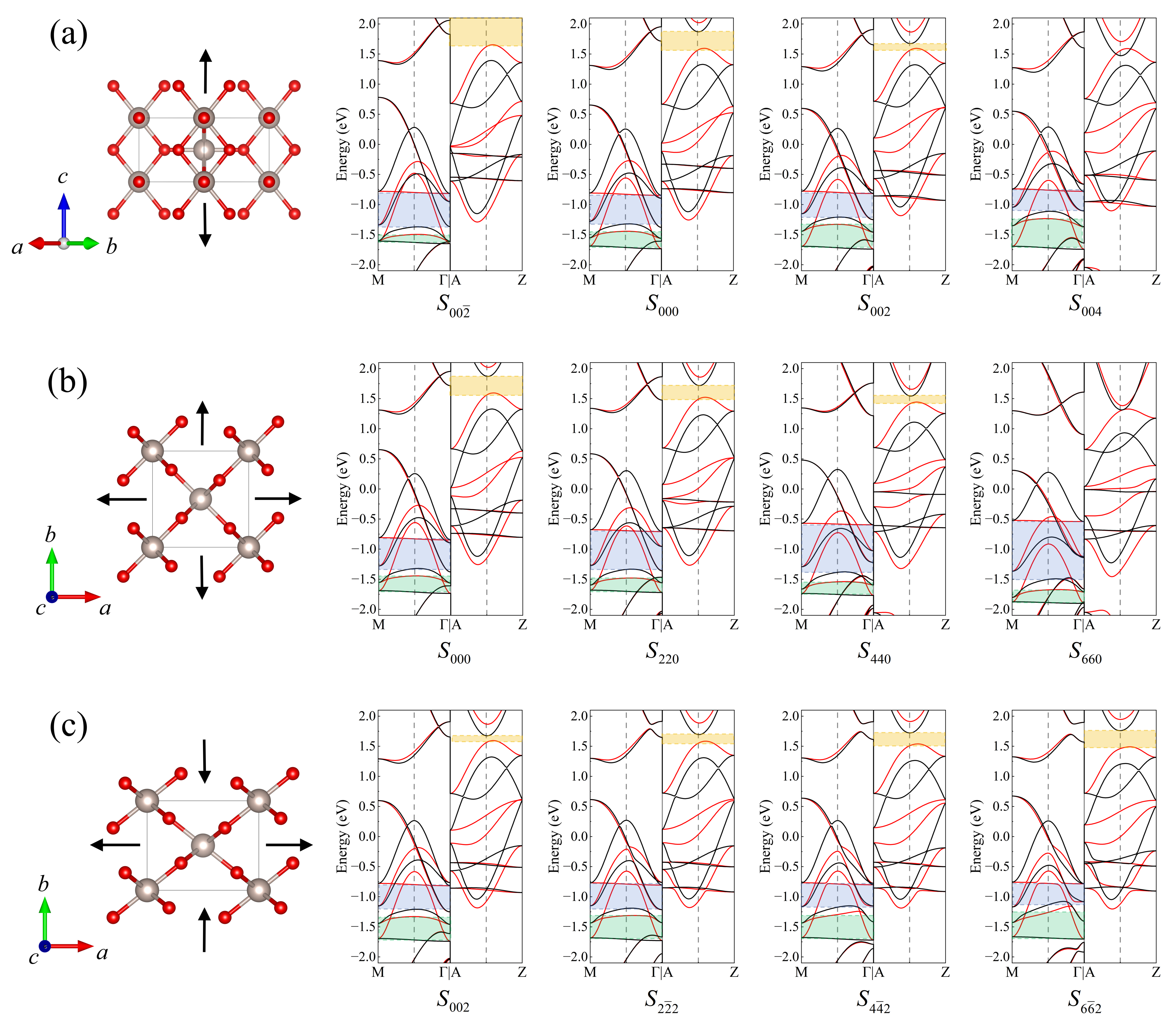}
	\caption{The electronic band structures of strained RuO$_2$ with (a): uniaxial strain along \textit{\textit{c}} axis, (b): isotropic biaxial tensile strain along \textit{\textit{a}} and \textit{\textit{b}} axes, and (c): equal amounts of tensile strain and compressive strain along \textit{\textit{a}} and \textit{\textit{b}} axes, respectively. The up spin and down spin bands are represented by the black and red lines, respectively. The featured spin-splitting in the middle of M-$\Gamma$ and A-Z between flat bands are marked by the blue, green or yellow boxes.
	}
	\label{fig5}
\end{figure}

Fig. \ref{fig5} shows the band structures of strained RuO$_2$. In principle, the spin-splitting widths of the occupied states are related to the magnitude of the spin magnetic moment, as the local spin polarization provides the effective exchange-splitting field. For uniaxial strain along the \textit{\textit{c}} axis (Fig. \ref{fig5}(a)), the spin-splitting widths between the flat bands vary considerably, with the spin magnetic moments of \textit{\textit{S}}\textsubscript{\textsubscript{00$\bar{2}$}}, \textit{\textit{S}}\textsubscript{\textsubscript{000}}, \textit{\textit{S}}\textsubscript{\textsubscript{002}}, and \textit{\textit{S}}\textsubscript{\textsubscript{004}} being 0.65 $\mu$\textsubscript{\textsubscript{B}}, 0.69 $\mu$\textsubscript{\textsubscript{B}}, 0.73 $\mu$\textsubscript{\textsubscript{B}}, and 0.77 $\mu$\textsubscript{\textsubscript{B}}, respectively. For isotropically biaxial strain along the \textit{\textit{a}} and \textit{\textit{b}} axes (Fig. \ref{fig5}(b)), the spin-splitting widths even vary significantly; the corresponding spin magnetic moments of \textit{\textit{S}}\textsubscript{\textsubscript{000}}, \textit{\textit{S}}\textsubscript{\textsubscript{220}}, \textit{\textit{S}}\textsubscript{\textsubscript{440}}, and \textit{\textit{S}}\textsubscript{\textsubscript{660}} are 0.69 $\mu$\textsubscript{\textsubscript{B}}, 0.80 $\mu$\textsubscript{\textsubscript{B}}, 0.92 $\mu$\textsubscript{\textsubscript{B}}, and 1.06 $\mu$\textsubscript{\textsubscript{B}}, respectively. Finally, for equal amounts of tensile and compressive strain along the \textit{\textit{a}} and \textit{\textit{b}} axes, respectively (Fig. 5(c)), the spin-splitting widths remain nearly unchanged, aside from some tilting of the flat bands. The spin magnetic moment is approximately unchanged 0.73 $\mu$\textsubscript{\textsubscript{B}} for \textit{\textit{S}}\textsubscript{\textsubscript{002}}, \textit{\textit{S}}\textsubscript{\textsubscript{2$\bar{2}$2}}, \textit{\textit{S}}\textsubscript{\textsubscript{4$\bar{4}$2}}, and \textit{\textit{S}}\textsubscript{\textsubscript{6$\bar{6}$2}}. Combined with the calculated results shown in Fig. 3, one can readily identify the covariance among crystal cell volume, spin splitting in bands, and spin magnetic moment in bulk RuO$_2$.

\begin{figure}[htbp]
	\centering
	\includegraphics[width=1.0\textwidth]{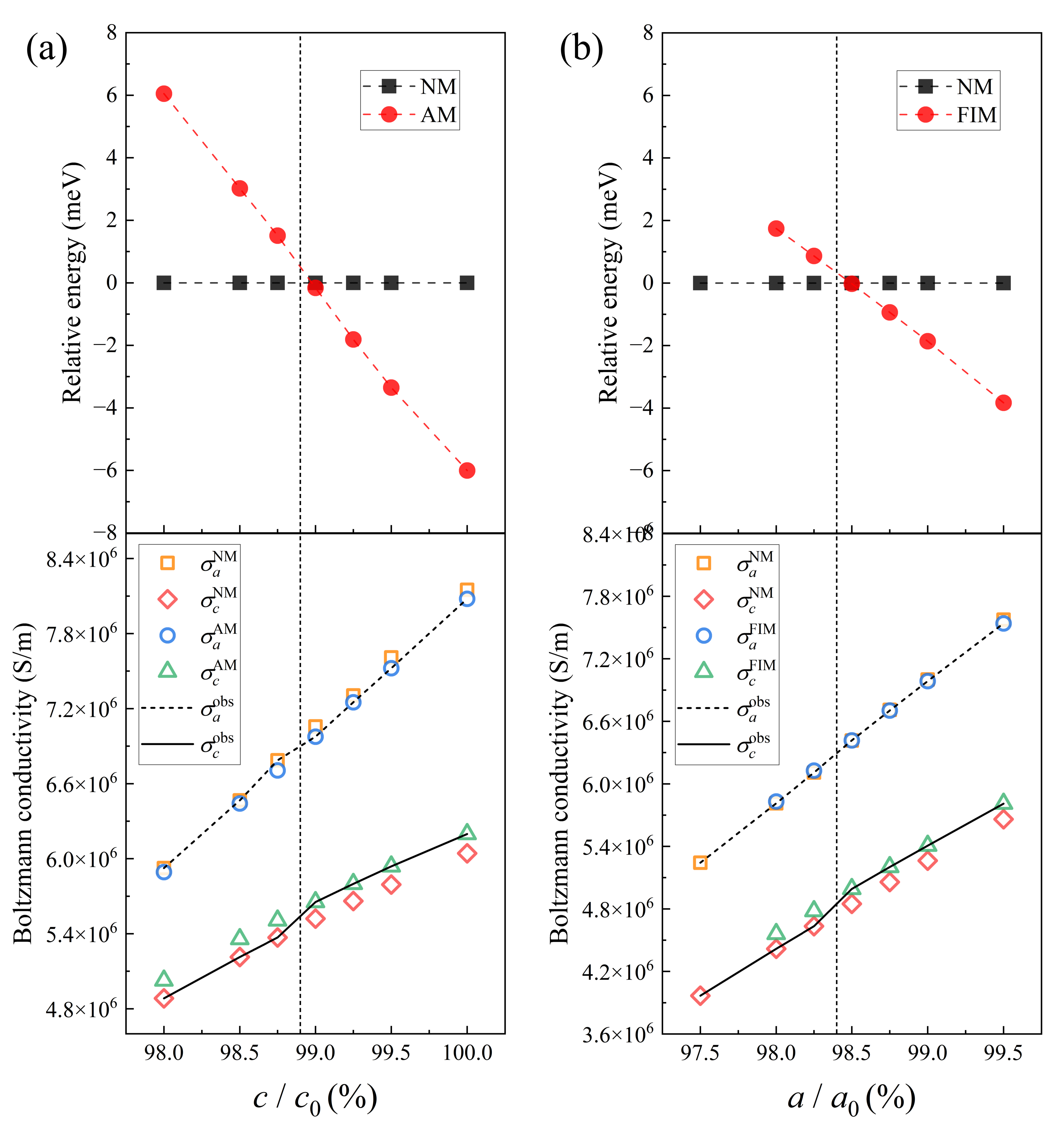}
	\caption{The relative energies (upper) and Boltzmann conductivities (lower) for the NM and AM/FIM states of bulk RuO$_2$ under continuous uniaxial strains along (a): $c$ axis and (b): $a$ axis. The $a$-axis- and $c$-axis- components of Boltzmann conductivities for the NM state are represented by the orange squares and red diamonds. The $a$-axis- and $c$-axis- components of Boltzmann conductivities for the AM/FIM states are represented by the blue circles and green triangles. The black dashed lines and black solid lines mark the actually observed longitudinal conductivity along $a$ axis and $c$ axis, respectively.  
	}
	\label{fig6}
\end{figure}

\subsection{\label{sec:level2}Discontinuity of the conductivity}
In this part, we discuss a straightforward way to experimentally verify the strain-induced NM-magnetic phase transition in bulk RuO$_2$. It has been suggested that the observation of the anomalous Hall effect (AHE) is a direct way to reflect the AM nature of a crystal sample \cite{Smejkal2022a,Reichlova2024}. Based on our prediction of NM-magnetic phase transitions in bulk RuO$_2$, we infer that the AHE can appear or disappear sharply at the transition points under continuous strains. Since the magnetic phase transition is of first order, the longitudinal conductivity may also undergo a discontinuous change due to the reconstruction of the Fermi surface, in addition to the change in the AHE. This dual verification of both transverse and longitudinal conductivity can provide strong experimental evidence of NM-magnetic transition in bulk RuO$_2$.

As a proof of principle, the Boltzmann transport conductivity of bulk RuO$_2$ under continuous uniaxial strains is simulated and shown in Fig. \ref{fig6}. The Boltzmann conductivity here is used to approximately reflect the experimentally observed longitudinal conductivity. The temperature is set to 300 K and the chemical potential is selected as a single value of 0. For continuous variation of the $c$ axis of bulk RuO$_2$ as shown in Fig. \ref{fig6}(a), the calculated conductivities of both the NM and AM states increase continuously. However, when crossing the NM-AM transition point, the actual observed conductivity exhibits a discontinuity. The discontinuous change of Boltzmann conductivity also shows anisotropy among different orientations of the crystal, with its components along $c$ and $a$ axes being positive and negative, respectively. For continuous variation of the $a$ axis of bulk RuO$_2$ as shown in Fig. \ref{fig6}(b), the conclusion is essentially the same except for a nearly zero component of discontinuous change in conductivity along the $a$ axis at the NM-FIM transition point. The anisotropy of the discontinuity in Boltzmann conductivity strongly depends on the anisotropic Fermi surfaces of the NM state and the AM/FIM states of bulk RuO$_2$.

\section{Conclusion}

In this work, the magnetic ground state of bulk rutile RuO$_2$ is investigated by DFT calculations. The spin magnetic moment in RuO$_2$ is shown to be sensitive to the 4\textit{\textit{d}}-electron correlation. Within the reasonable range of effective Coulomb parameter \textit{\textit{U}}\textsubscript{\textsubscript{eff}} for RuO$_2$, there exists multiple AM phases. By applying various strains to the RuO$_2$ bulk, the ground state can undergo transitions between the NM state and magnetic ordered states. Moreover, the magnitude of spin magnetic moment and spin-splitting widths in the band structure can be continuously tuned through strain engineering. From an alternative perspective that is independent of surface effects and defects in RuO$_2$ crystalline samples, our proposed correlation-sensitive and strain-dependent magnetism in rutile RuO$_2$ bulk provides additional insights to resolve the experimental controversy regarding its magnetic ground state.

According to the calculated magnetic phase diagram shown in Fig. \ref{fig3}, whether the magnetic moments in bulk rutile RuO$_2$ occur or not mainly depends on the cell volume, rather than any specific strain modes reported \cite{Forte2026,Zhang2025b,Alaei2026,Jiang2026,Jeong2026a,Wickramaratne2026}. This property can expand and simplify the experimental methods for tuning the magnetism of RuO$_2$ through strain engineering. Compared to the epitaxial strain for RuO$_2$ thin films, we suggest that any imposed mechanical-strain mode that significantly changes the volume of RuO$_2$ crystal could switch on/off the spin polarization in bulk RuO$_2$ and the spin splitting in its band structure. 

\begin{acknowledgments}
This work is supported by the National Natural Science Foundation of China (Grant No. 12274028, No. 12234003, No. W2511003, and No. 12274027), the National Key R$\&$D Program of China (2022YFA1402603). We thank Prof. Run-Wu Zhang , Prof. Wayne Zheng, and Dr. Yichen Liu from School of physics, Beijing Institute of Technology for their useful discussions with us.
\end{acknowledgments}

\nocite{*}
\bibliography{references}

\end{document}